\documentclass[preprintnumbers,showpacs,amsmath,amssymb,floatfix,9pt,prd,onecolumn,
superscriptaddress,nofootinbib]{revtex4}
\usepackage{graphicx}
\usepackage{latexsym}
\usepackage{epsfig}
\usepackage{amssymb}
\usepackage[utf8]{inputenc}
\begin{document}
\title{Tolman IV fluid sphere in $f(R,\,T)$ gravity }

\author{Piyali Bhar \footnote{Corresponding author}}
\email{piyalibhar90@gmail.com , piyalibhar@associates.iucaa.in}
\affiliation{Department of
Mathematics,Government General Degree College, Singur, Hooghly, West Bengal 712409,
India}

\author{Pramit Rej}
\email{pramitrej@gmail.com} \affiliation{Department of
Mathematics, Sarat Centenary College, Dhaniakhali, Hooghly, West Bengal 712 302, India}

\author{M. Zubair}
\email{mzubairkk@gmail.com; drmzubair@cuilahore.edu.pk}\affiliation{Department
of Mathematics, COMSATS University Islamabad, Lahore Campus, Lahore,
Pakistan}

\begin{abstract}
In this article, we studied the behavior of relativistic spherical objects considering Tolman IV spacetime in modified $f(R,\,T)$ gravity for the uncharged perfect fluid matter. We have chosen the matter Lagrangian as $\mathcal{L}_m=-p$ to develop our present model. In particular, for this investigation we have reported for the compact object LMC $X-4$ [Mass=$(1.04 \pm 0.09)M_{\odot}$; Radius= $8.301_{-0.2}^{+0.2}$ Km] in our paper.
The effect of the coupling parameter $\beta$ on the local matter distribution of compact stars has been investigated in this paper. It can be seen that with greater values of $\beta$, the sound speed and adiabatic index are higher. On contrary, the mass function takes lower value for higher values of $\beta$. Our obtained solution does not admit singularities in the matter density, pressure and metric functions.
According to our graphical analysis, this new stellar model satisfies all physical requirements anticipated in a realistic star.
\end{abstract}



\maketitle

\section{Introduction}

General relativity (GR) has proved itself a tremendously successful theory, which provides fruitful information about the evolution and
hidden secrets of the universe. However, it is exposed to some serious challenges in the the presence of dark elements, i.e., dark matter and dark energy.
According to the predictions of GR, a universe dominated by matter or radiation accelerates in a negative direction due to the gravitational
pull. Nevertheless, current astronomical observations uncover this scenario and confirms the accelerated expansion of the universe \cite{ob1,ob2,ob3,ob4,ob5,ob6,ob7,ob8,ob9,ob10,ob11}.
In this situation, it is essential to modify the framework of GR, so that cosmological and astrophysical phenomena caused by dark components can be best
explained. In this connection, $f(R)$ theory is an important milestone \cite{fr1,fr2}, which is based on the modification in the geometrical sector.
It modifies the standard Einstein-Hilbert (EH) action by inserting an arbitrary function $f(R)$ instead of the Ricci scalar $R$. After the innovational
work on cosmological inflation in the realm of $f(R)$ gravity \cite{fr3}, it has become an active research arena. Qadir and his co-researchers \cite{fr4}
strengthened the idea of modified relativistic dynamics and pointed out the role of this modification in order to resolve the issues
related to dark matter and dark energy.

Following the idea of introducing alteration in EH action, Harko et al. \cite{harko11} presented $f(R, T)$ theory in which trace of stress energy
tensor $T$ has been introduced as a new ingredient. The appearance of this new ingredient is associated with the quantum effects or imperfect fluids.
This modified theory consists on minimal matter coupling between the matter and gravitational sectors, which facilitates the study of new gravitational aspects.
In literature, there has been a huge amount of research work on different aspects of $f(R, T)$ theory. It has been tested
in different dimensions which include cosmology \cite{frt5,frt6,frt7,frt8,frt9}, thermodynamics \cite{frt10,frt11},
gravitational waves \cite{frt12} and astrophysics of stellar systems \cite{frt13,frt14,f1,f2,f3,f4,f5}, and it has presented valuable
contributions to the different cosmological and astrophysical issues \cite{frt15,frt16,frt17,frt18,frt19,frt20,frt21}.
Some of them present collapsing of the non-static spherically symmetric
objects with anisotropic fluid profile \cite{frt22} and wormhole solutions with the static spherically symmetric geometry \cite{frt23,frt24}.
Furthermore, perturbation techniques were employed for the study of self-gravitating systems with cylindrically symmetric geometry \cite{frt25}.
Houndjo explored  implications of $f(R, T)$ theory on gravitational lensing \cite{frt26}.
Baffou and his collaborators \cite{frt27} applied perturbation approach and worked out
some cosmic limits on power-law models and de-Sitter space-time. Recently, great efforts have been
made to explore the existence of collapsing structures
in the background of this modified theory \cite{frt28, frt29}. Moreover, it
can be considered as the most reliable one for the exploration of cosmological aspects.
In last few years, this gravity theory has been considered in the study of compact structures.
Moraes and his co-researchers \cite{frt30} worked out modified Tolman-Oppenheimer-Volkoff (TOV) equation,
which depicts equilibrium condition of the compact structures.

The study of different compact structures and self-gravitating objects plays a fundamental role in revealing different aspects of the universe.
One of the main evolutionary phases of the celestial structures is the phenomenon of gravitational collapse near their death which
is responsible for the formation of compact objects. These compact objects appear as end-points during the evolution of
an ordinary celestial system, which can be treated as an ideal source for exploring the features and characteristics of highly
dense matter distributions. In this regard, different compact objects with extremely high energy densities
has been found \cite{c1}, which are usually identified as pulsars and spinning stars with strong magnetic fields.
At theoretical levels, our understanding of compact objects has its origin in
the Fermi-Dirac statistics due to the high degeneracy pressure of these objects
which avoids the exposure of the stellar systems from gravitational collapsing \cite{c2}.
Chandrasekhar \cite{c3,c4} identified white dwarfs as compact structures which secure their existence by
a degenerate gas of electrons, and showed that their stability depends on their size which
should be around $1.4$ times of the solar mass for the greatest one.

In different aspects, the compact structures have gained a lot of interest, however there is significantly more to be probed and explored about these stellar structures. These compact structures are assumed to be highly dense due to their large masses and small radii, which may be described in the framework of GR and modified theories effectively \cite{c5, c6}. The spherically symmetric sources with isotropic fluid distribution are assumed to be the simplest or the most most trivial because different observational data sets corroborates the existence of isotropy in fluid configurations. A theoretical algorithm was suggested to generate any number of pressure and density profiles for isotropic distributions without evaluation of the integrals \cite{c7}. In \cite{hansraj},
authors discussed the embedding class charged isotropic stars and found that conformally flat charged isotropic stars of embedding class one do not exist. It is shown that Finch–Skea type model is the only choice if spacetime accepts conformal symmetries. To analyze the geometry of compact stellar structures, the implications of different modified gravity theories are well-known. Abbas \cite{c8} analyzed equilibrium condition for compact stellar structures and explored their physical features in the framework of modified Gauss-Bonnet gravity. Zubair and Abbas \cite{c9} considered $f(R)$ theory and presented their analysis on the geometry of compact stellar systems with anisotropic matter configuration. The research in the arena of compact structures mainly concerns with isotropic, anisotropic and charged fluids in the background of $f(R, T)$ gravity
and provides interesting information about the features and evolution of these structures \cite{c10,c11,c12,c13,c14,c15}.

Motivated by the previous studies, we are interested to explore isotropic Tolman IV model in the background of $f(R,T)$ theory of gravity. The physical attributes of obtained model are examined for well-known compact star model LMC X-4. The paper is arranged in the following order of sections. In next section we explain about basic equations of $f(R,T)$ theory and proposed model is obtained in section \ref{sec3}. In section \ref{sec4}, we match our interior spacetime to the exterior spacetime at the boundary. Section \ref{sec5} explains the physical analysis of the obtained model subject to different values of coupling parameter $\beta$. Final section \ref{sec6} concludes the whole work by focusing on major findings.

\section{ Basic field Equations}\label{sec2}
We consider the most general action for $f(R,\,T)$ gravity to study the stellar configurations of compact stars in modified $f(R,\,T)$ gravity proposed by Harko {\em et al.} \cite{harko11} as,
\begin{eqnarray}\label{action}
S&=&\int \left( \frac{1}{16 \pi} f(R,T) +  \mathcal{L}_m\right)\sqrt{-g} d^4 x,
\end{eqnarray}
where $R$ is the Ricci scalar and $T$ is the trace of the energy-momentum tensor $T_{\mu \nu}$, $\mathcal{L}_m$ represents the lagrangian matter density and $g = det(g_{\mu \nu}$). The modified field equations are determined by varying the action (\ref{action}) with regard to the metric tensor $g_{\mu\nu}$ which yields,
\begin{eqnarray}\label{frt}
f_R(R,T)R_{\mu \nu}-\frac{1}{2}f(R,T)g_{\mu \nu}+(g_{\mu \nu }\Box-\nabla_{\mu}\nabla_{\nu})f_R(R,T)&=&8\pi T_{\mu \nu}-f_T(R,T)T_{\mu \nu}\nonumber\\&&-f_T(R,T)\Theta_{\mu \nu},
\end{eqnarray}
where the subscript $R$ and $T$ in $f_R(R,\,T)$ and $f_T(R,\,T)$ represents the partial derivative with respect to $R$ and $T$ respectively. $\nabla_{\nu}$ denotes the covariant derivative
associated with the Levi-Civita connection of $g_{\mu \nu}$, $\Theta_{\mu \nu}=g^{\alpha \beta}\frac{\delta T_{\alpha \beta}}{\delta g^{\mu \nu}}$ and
$\Box \equiv \frac{1}{\sqrt{-g}}\partial_{\mu}(\sqrt{-g}g^{\mu \nu}\partial_{\nu})$ represents the D'Alambert operator.\\
The compact star model with perfect fluid is used in this investigation. As a result, the stress-energy tensor of the matter is given by,
\begin{eqnarray}
T_{\mu \nu}&=&(p+\rho)u_{\mu}u_{\nu}-p g_{\mu \nu},
\end{eqnarray}
where $\rho$ is the matter density, $p$ is the isotropic pressure in modified gravity, $u^{\mu}=e^{\frac{\nu}{2}}\delta_{\alpha}^0$ is the fluid four velocity satisfies the equations $u^{\mu}u_{\mu}=1$ and $u^{\mu}\nabla_{\nu}u_{\mu}=0$. The stress-energy tensor of matter proposed by Landau and Lifshitz \cite{landau}, is defined as,
\begin{eqnarray}\label{tmu1}
T_{\mu \nu}&=&-\frac{2}{\sqrt{-g}}\frac{\delta \sqrt{-g}\mathcal{L}_m}{\delta \sqrt{g_{\mu \nu}}},
\end{eqnarray}
with trace $T=g^{\mu \nu}T_{\mu \nu}$. Now eqn. (\ref{tmu1}) turns out the following form if the Lagrangian density $\mathcal{L}_m$ depends only on $g_{\mu \nu}$, not on its derivatives,
\begin{eqnarray}
T_{\mu \nu}&=& g_{\mu \nu}\mathcal{L}_m-2\frac{\partial \mathcal{L}_m}{\partial g_{\mu \nu}}.
\end{eqnarray}
For our present study we choose the matter Lagrangian as $\mathcal{L}_m=-p$ following Harko et al. \cite{harko11} and the expression of $\Theta_{\mu \nu}$ becomes, $\Theta_{\mu \nu}=-2T_{\mu \nu}-pg_{\mu\nu}.$\\
By taking the covariant divergence of (\ref{frt}) (see refs. \cite{harko11,hi}), we get,
\begin{eqnarray}\label{conservation}
\nabla^{\mu}T_{\mu \nu}&=&\frac{f_T(R,T)}{8\pi-f_T(R,T)}\left[(T_{\mu \nu}+\Theta_{\mu \nu})\nabla^{\mu}\ln f_T(R,T)+\nabla^{\mu}\Theta_{\mu \nu}\right],
\end{eqnarray}
which is the divergence of the stress-energy tensor $T_{\mu \nu}$. From eqn. (\ref{frt}) the field equations of $f(R)$ gravity can be obtained when $f(R,T)=f(R)$.
The above equation indicates that $\nabla^{\mu}T_{\mu \nu} \neq 0$ if $f_T(R,T) \neq 0$, that is why the system will not be conserved like Einstein gravity.\par

\section{Interior Spacetime and the realistic viable $f(R,\,T)$ gravity models}\label{sec3}
In this section we will describe the model of compact stars by using a realistic $f(R,\,T)$ gravity model. Let us assume a separable functional form for $f(R,\,T)$ given by,
\begin{eqnarray}
f(R,\,T) = f_1(R)+f_2(T ),
\end{eqnarray}
in relativistic structures to discuss the coupling effects of matter and curvature components in $f(R,\,T)$ gravity, where $f_1(R)$ and $f_2(T)$ being arbitrary functions of $R$ and $T$ respectively. Several viable models in $f(R,\,T)$ gravity can be generated by combining different forms of $f_1(R)$ with a linear combination of $f_2(T)$. In our present model, we consider $f_1(R)=R$ and $f_2(T)=2\beta T$, i.e., the expression of $f(R,\,T)$ becomes
\begin{eqnarray}\label{e}
f(R,T)&=& R+2 \beta T,
\end{eqnarray}
$\beta$ is arbitrary constant to be estimated depending on several physical requirements.\par
The static and spherically symmetric line element in curvature coordinates $(t,\,r,\,\theta,\,\phi)$ is given by,
\begin{equation}\label{line}
ds^{2}=-e^{\nu}dt^{2}+e^{\lambda}dr^{2}+r^{2}\left(\sin^{2}\theta d\phi^{2}+d\theta^{^2}\right),
\end{equation}
the metric co-efficients $\nu$ and $\lambda$ depend only on `r', i.e., they are purely radial.

The field equations in $f(R,T)$ gravity is given by,
\begin{eqnarray}
G_{\mu \nu}&=&8\pi T_{\mu \nu}^{\text{eff}},
\end{eqnarray}
where $G_{\mu \nu}$ is the Einstein tensor and
\begin{eqnarray}
T_{\mu \nu}^{\text{eff}}&=& T_{\mu \nu}+\frac{\beta}{8\pi}T g_{\mu \nu}+\frac{\beta}{4\pi}(T_{\mu \nu}+p g_{\mu \nu}).
\end{eqnarray}
For the line element (\ref{line}), the field equations in modified gravity can be written as,
\begin{eqnarray}
8\pi\rho +\beta (3 \rho-p)&=&\frac{\lambda'}{r}e^{-\lambda}+\frac{1}{r^{2}}(1-e^{-\lambda}),\label{f1}\\
8 \pi p-\beta (\rho-3p)&=& \frac{1}{r^{2}}(e^{-\lambda}-1)+\frac{\nu'}{r}e^{-\lambda},\label{f2} \\
8 \pi p-\beta (\rho-3p)&=&\frac{1}{4}e^{-\lambda}\left[2\nu''+\nu'^2-\lambda'\nu'+\frac{2}{r}(\nu'-\lambda')\right]. \label{f3}
\end{eqnarray}
We denote $\rho^{\text{eff}}$ and $p^{\text{eff}}$ by,
\begin{eqnarray}
\rho^{\text{eff}}&=& \rho+\frac{\beta}{8\pi}(3 \rho-p),\label{r1}\\
p^{\text{eff}}&=& p-\frac{\beta}{8\pi}(\rho-3p),\label{r2}
\end{eqnarray}
where $\rho^{\text{eff}}$ and $p^{\text{eff}}$ respectively denote the density and pressure in Einstein gravity and
the `prime' denotes differentiation with respect to `r'. \\

Equations (\ref{f1})-(\ref{f3}) are highly non-linear in nature. To solve these equations, for our present model, we choose the expression for $e^{\lambda}$ as,
\begin{eqnarray}\label{elambda}
e^{\lambda}&=&\frac{1 + 2ar^2}{(1 - br^2)(1 + ar^2)},
\end{eqnarray}
where $a,\,b$ are arbitrary constants of dimension km$^{-2}$. This metric potential was initially proposed by Tolman \cite{tol}. Tolman IV metric potential was used earlier by several researchers to model compact objects \cite{t1,t2,t3,t4,t5,t6,t7,t8}.\par
Using the expression of $e^{\lambda}$ from eqn. (\ref{elambda}) and eqn. (\ref{f1}), we obtain,
\begin{eqnarray}
\rho^{\text{eff}}&=&\frac{3 b + a \{3 + (2 a + 7 b) r^2 + 6 a b r^4\}}{8\pi(1 + 2 a r^2)^2},
\end{eqnarray}
Now the eqns. (\ref{f2}) and (\ref{f3}) together imply,
\begin{eqnarray}\label{c}
r^2(2\nu''+\nu'^2-\nu'\lambda')-2r(\nu'+\lambda')+4(e^{\lambda}-1)=0.
\end{eqnarray}
Using the expression of $e^{\lambda}$ into eqn. (\ref{c}), we obtain the expression of the another metric potential as,
\begin{eqnarray}
e^{\nu}&=&B^2(1 + ar^2),
\end{eqnarray}
where $B$ is a constant of integration and it is a dimensionless quantity. The expression for $p^{\text{eff}}$ is thus obtained as,
\begin{eqnarray}
p^{\text{eff}}&=&\frac{a - b - 3 a b r^2}{8\pi(1 + 2 a r^2)}.
\end{eqnarray}
Using the expression of $p^{\text{eff}}$ and $\rho^{\text{eff}}$, from eqns. (\ref{r1}) and (\ref{r2}), we obtain the expression of matter density and pressure $\rho,\,p$ in modified gravity as,
\begin{eqnarray}
\rho&=&\frac{5 a \beta + 4 b \beta + 12 a \pi + 12 b \pi +
 4 a (a \beta + 2 b \beta + 2 a \pi + 7 b \pi) r^2 +
 6 a^2 b (\beta + 4 \pi) r^4}{4 (\beta + 2 \pi) (\beta + 4 \pi) (1 +
   2 a r^2)^2},\label{p1}\\
p&=&\frac{1}{8} \Big[-\frac{3 b}{
    \beta + 2 \pi} + \frac{(2 a + b) \beta}{(\beta + 2 \pi) (\beta +
      4 \pi) (1 + 2 a r^2)^2} + \frac{
   2 (2 a + b)}{(\beta + 4 \pi) (1 + 2 a r^2)}\Big]\label{p2}
\end{eqnarray}
\begin{figure}[htbp]
    \centering
        \includegraphics[scale=.55]{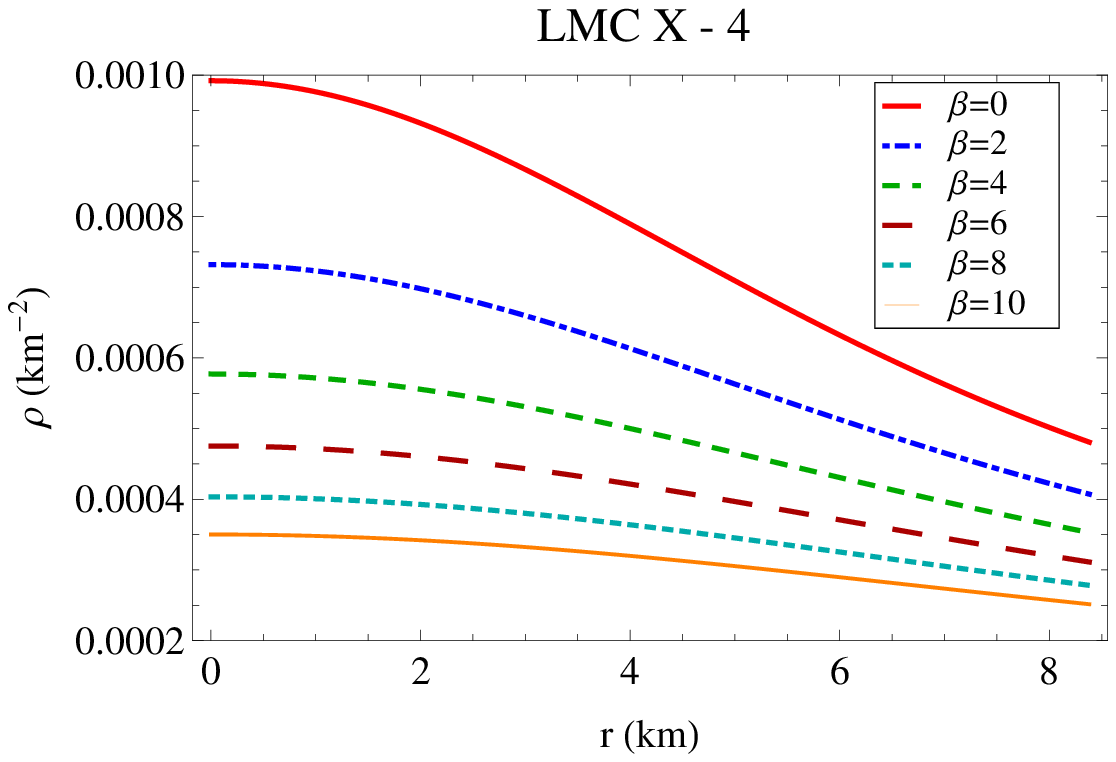}
        \includegraphics[scale=.55]{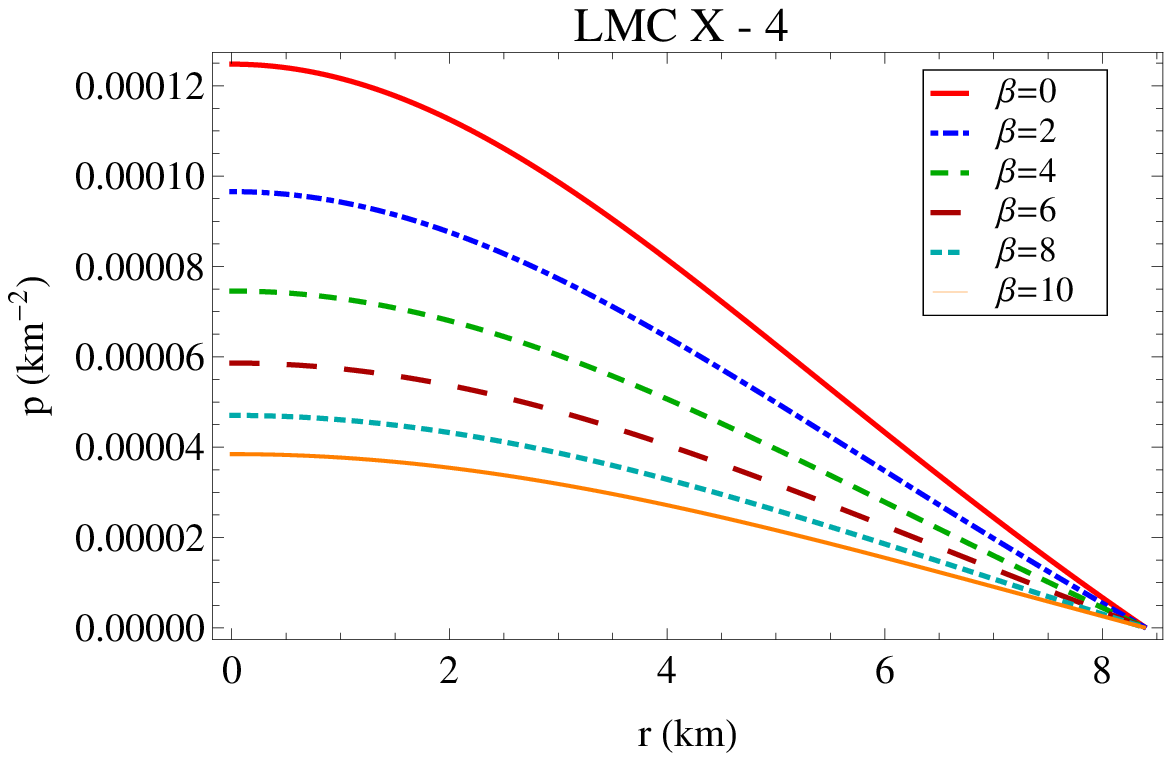}
       \caption{(left) Matter density and (right) pressure are plotted against radius for different values of the coupling constant mentioned in the figure. \label{pp}}
\end{figure}
The profiles of pressure and density are shown in Fig.~\ref{pp}.\\
Simply taking the differentiation of the expressions of $\rho$ and $p$ given in eqns.(\ref{p1})-(\ref{p2}) yield the pressure and density gradient as,
\begin{eqnarray*}
\rho'&=& -\frac{2 a (2 a + b) r \{2 \beta + 5 \pi + a (\beta + 2 \pi) r^2\}}{(\beta +
    2 \pi) (\beta + 4 \pi) (1 + 2 a r^2)^3},\\
         p'&=&-\frac{2 a (2 a + b) r \{\beta + \pi + a (\beta + 2 \pi) r^2\}}{(\beta +
    2 \pi) (\beta + 4 \pi) (1 + 2 a r^2)^3}.
\end{eqnarray*}
\begin{figure}[htbp]
    \centering
        \includegraphics[scale=.55]{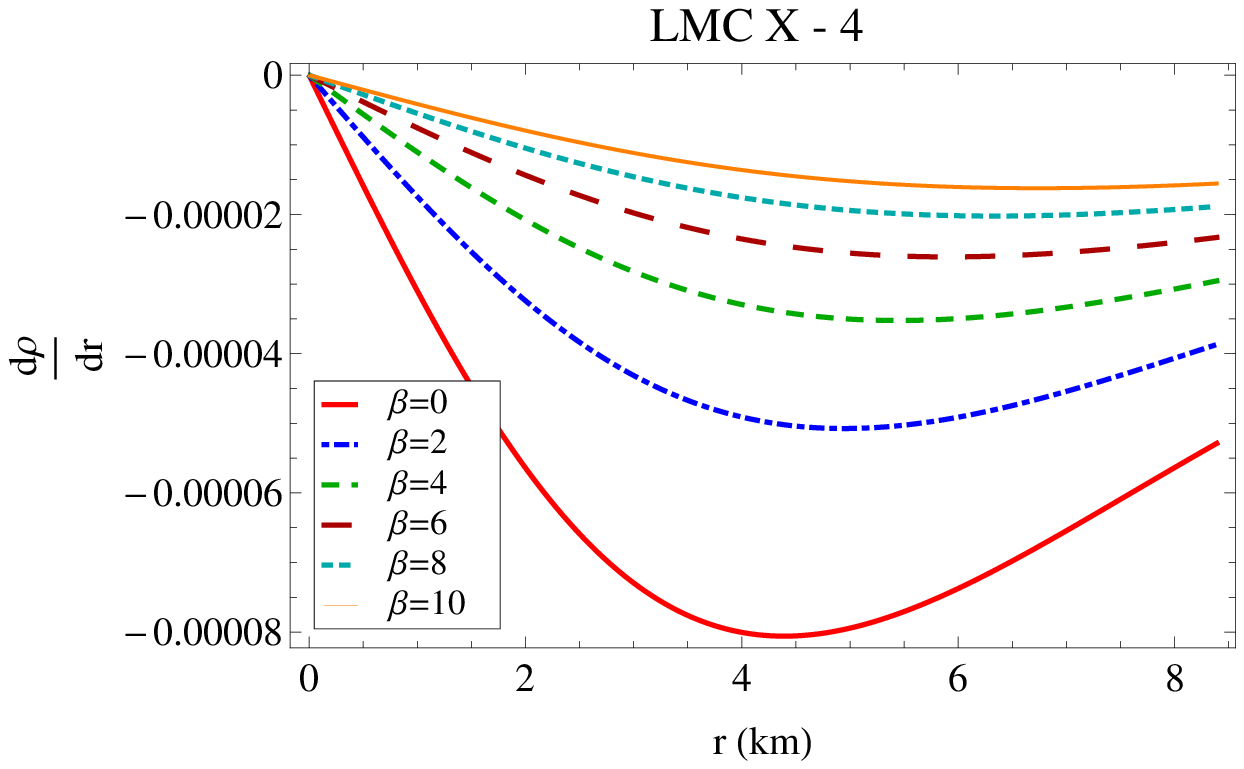}
        \includegraphics[scale=.55]{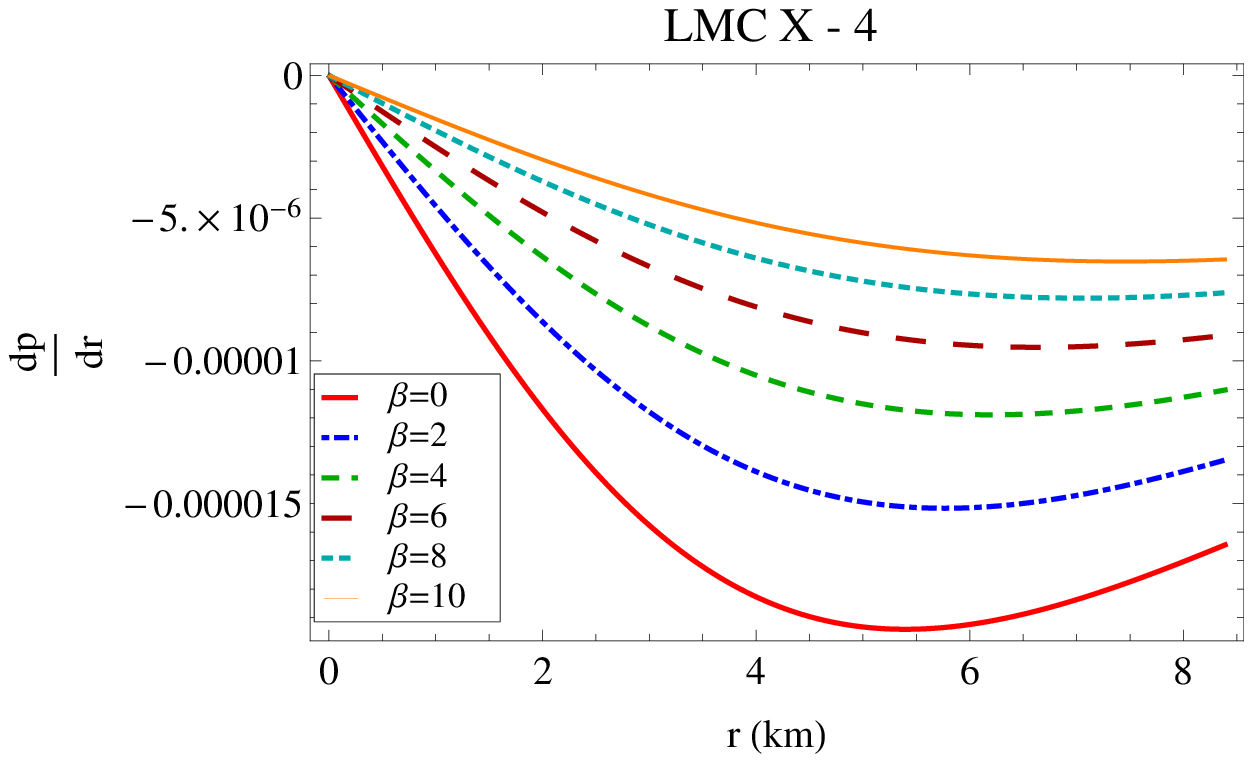}
       \caption{The pressure and density gradients are shown against `r'.}\label{grad}
\end{figure}

The behavior of pressure and density gradient are shown in Fig.~\ref{grad} for different values of $\beta$.\\
Now we investigate the stability criteria for a physically realistic anisotropic stellar compact object using graphical representation by employing numerical values for several unknown constants. We use the causality condition to demonstrate this criteria. The square of the speed of sound $V^2$ in the entire region of the fluid sphere must follow the bound $0<V^2<1$ to satisfy causality condition. For our present model the square of the sound velocity is obtained as,
\begin{eqnarray}
V^2&=&\frac{dp}{d\rho}=\left(\frac{dp}{dr}\right)/\left(\frac{d\rho}{dr}\right)=\frac{\beta + \pi + a (\beta + 2 \pi) r^2}{
2 \beta + 5 \pi + a (\beta + 2 \pi) r^2},
\end{eqnarray}
\begin{figure}[htbp]
    \centering
        \includegraphics[scale=.55]{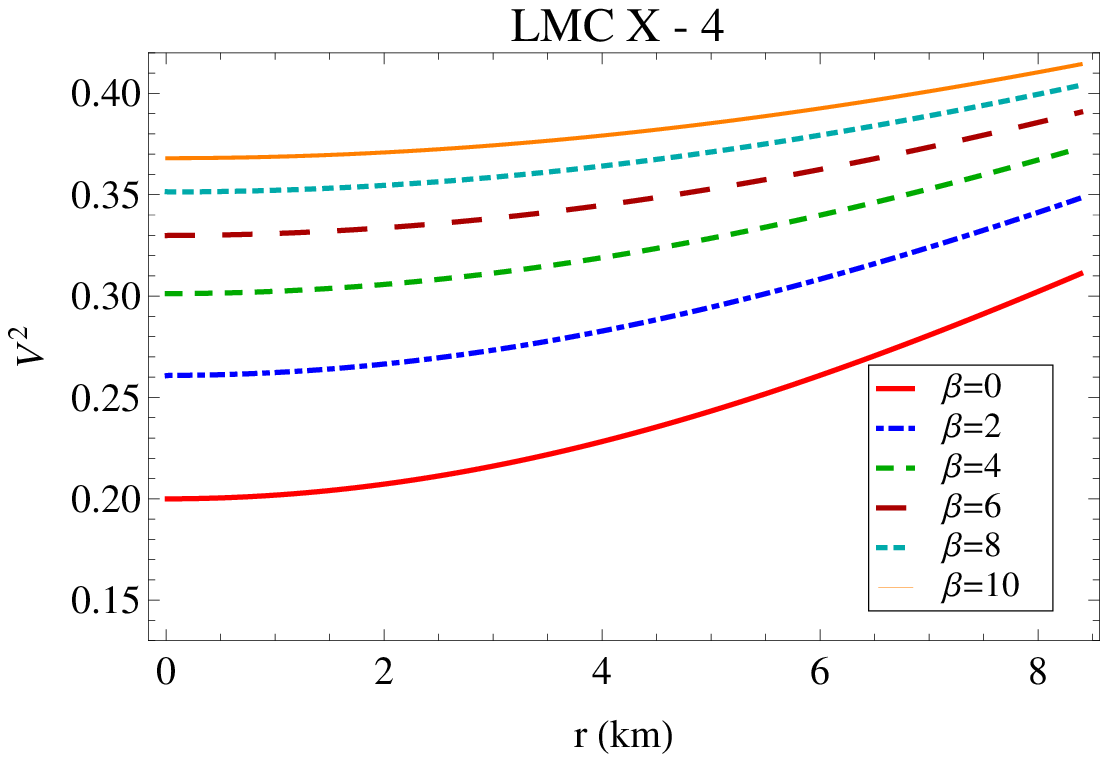}
        \includegraphics[scale=.55]{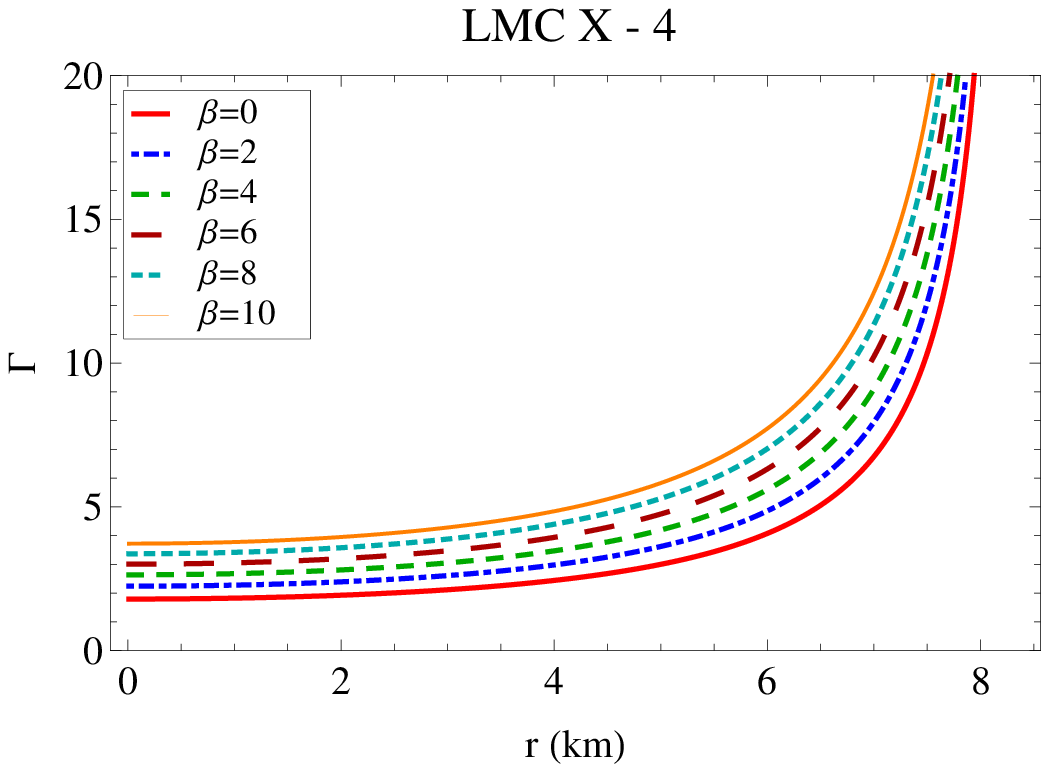}
       \caption{(left) The square of the sound velocity and (right) relativistic adiabatic index are plotted against radius inside the stellar interior.\label{sv}}
\end{figure}
In Fig.~\ref{sv}, the graphical nature of the causality condition for the compact object LMC X-4 is examined for different values of $\beta$, from which it is clear that the square of the sound speed lies within the predicted range throughout the fluid sphere.\\
Energy conditions are a set of physical properties that can be used to explore the presence of ordinary and exotic matter inside a star formation. The validity of the second law of black hole thermodynamics and the Hawking-Penrose singularity theorems can be easily tested using the energy conditions \cite{pen}. These conditions of energy are referred to as null, weak, strong and dominant energy conditions, symbolized respectively by NEC, WEC, SEC and DEC. All energy conditions for our current model are met if the following inequalities are hold :
    \begin{eqnarray*}
 \text{NEC}:~\rho+p\geq 0,\, \text{WEC}:~\rho+p\geq 0,~ \rho \geq 0,\, \text{SEC}:~\rho+p \geq 0, \rho+ 3p \geq 0,\,\text{DEC}:~\rho-p\geq 0,~ \rho \geq 0.
\end{eqnarray*}
To check the aforementioned energy conditions, we shall require the following expressions.
\begin{eqnarray}
\rho+p&=&\frac{(2 a + b) (1 + a r^2)}{(\beta + 4 \pi) (1 + 2 a r^2)^2},\\
\rho+3p&=&\frac{7 a \beta + 2 b \beta + 12 a \pi +
 2 a (4 a (\beta + 2 \pi) - b (\beta + 8 \pi)) r^2 -
 6 a^2 b (\beta + 4 \pi) r^4}{2 (\beta + 2 \pi) (\beta + 4 \pi) (1 +
   2 a r^2)^2},\\
   \rho-p&=&\frac{1}{4 (\beta + 2 \pi)}\left[3 b + \frac{2 a + b}{(1 + 2 a r^2)^2}\right].
\end{eqnarray}

\begin{figure}[htbp]
    \centering
        \includegraphics[scale=.55]{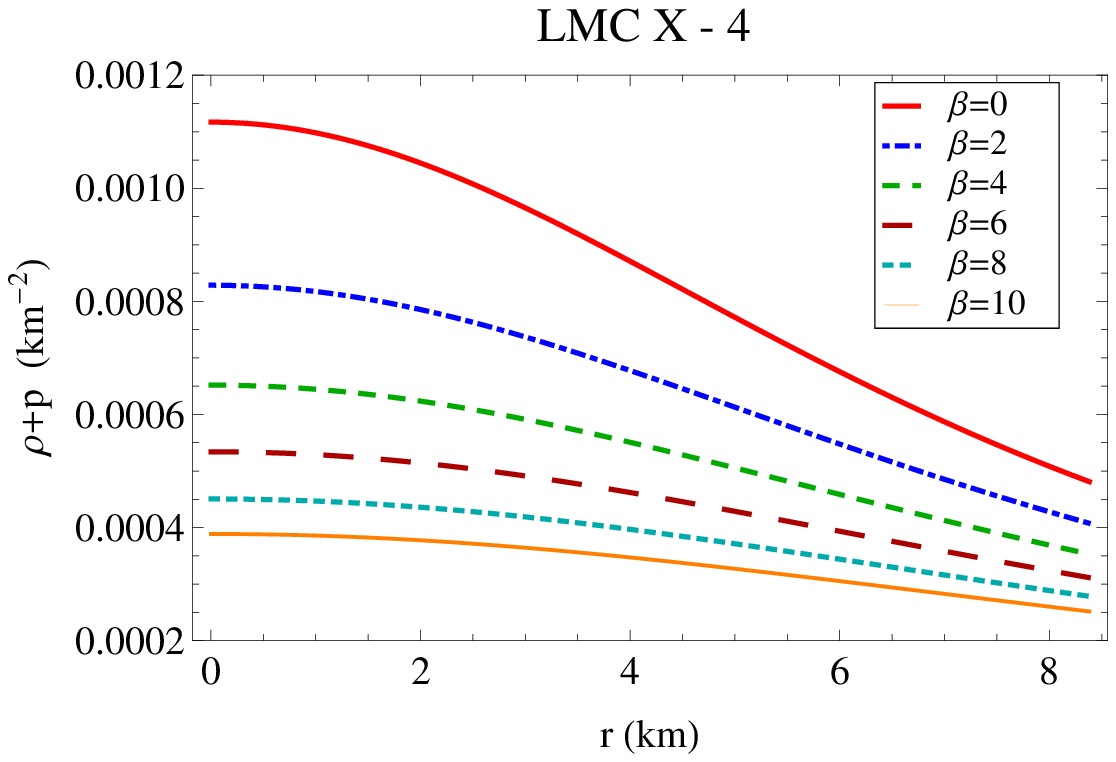}
        \includegraphics[scale=.55]{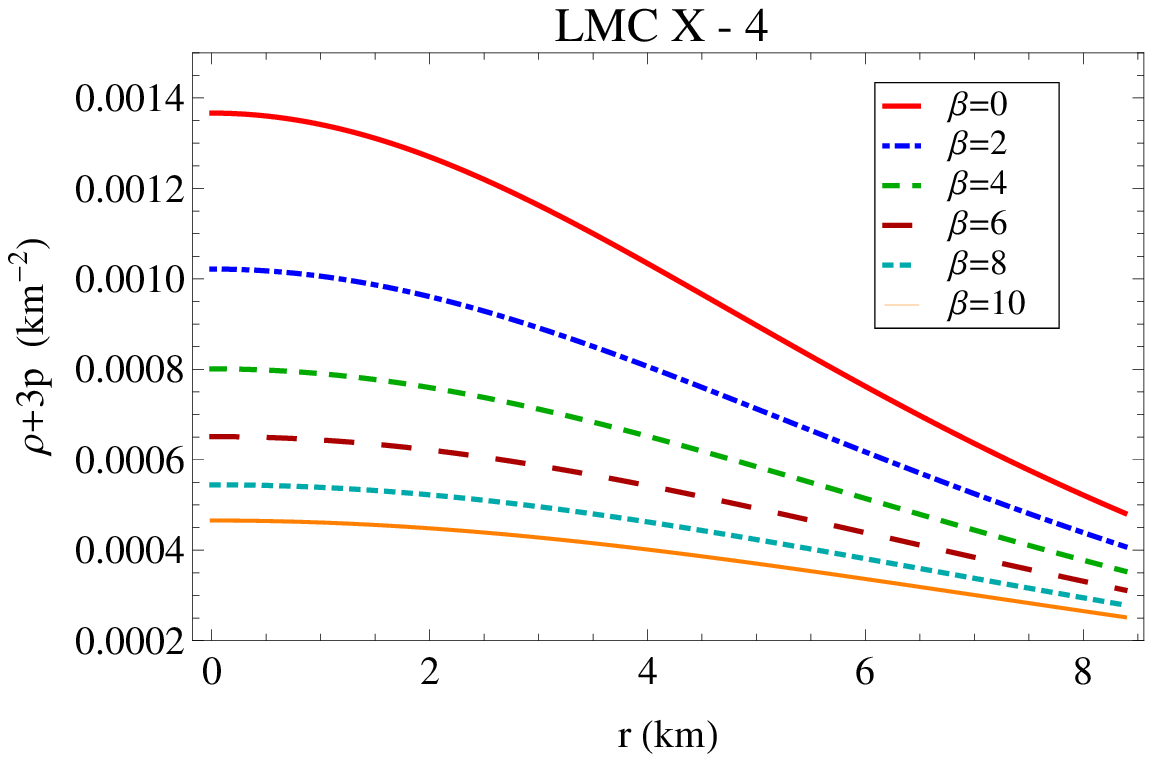}
        \includegraphics[scale=.55]{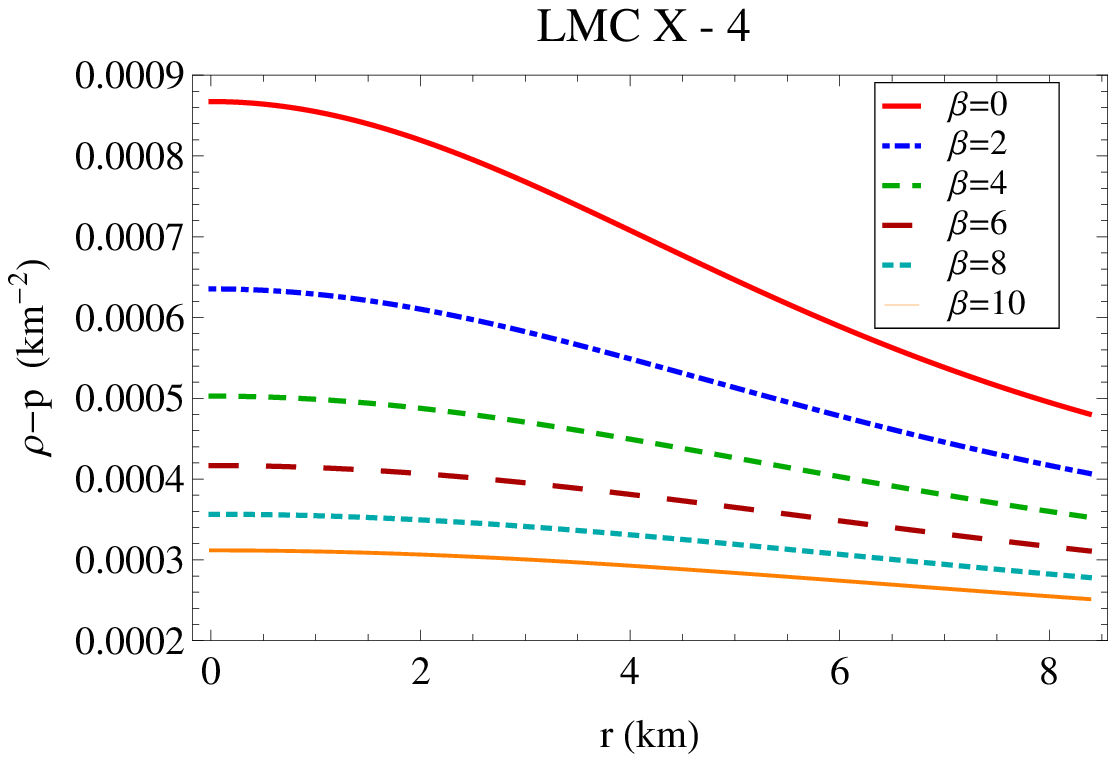}
       \caption{All the energy conditions are plotted against radius for different values of the coupling constant mentioned in the figure.\label{ec}}
\end{figure}

All of the energy conditions for our chosen $f(R,\,T)$ model have been satisfied, as shown graphically in Fig.~\ref{ec}.

\section{Exterior spacetime and boundary condition}\label{sec4}
We now match our interior spacetime to the exterior Schwarzschild line element at the boundary $r=R$. The exterior line element is given by,
\begin{eqnarray}
ds_+^{2}&=&-\left(1-\frac{2M}{r}\right)dt^{2}+\left(1-\frac{2M}{r}\right)^{-1}dr^{2}+r^{2}\left(d\theta^{2}+\sin^{2}\theta d\phi^{2}\right),
\end{eqnarray}
corresponding to the interior spacetime ,
 \begin{eqnarray}
ds_{-}^2 & = &-B^2\left(1 + ar^2\right) dt^2+ \frac{1 + 2ar^2}{(1 - br^2)(1 + ar^2)} dr^2+r^2(d\theta^2+\sin^2 \theta d\phi^2).
\end{eqnarray}
The continuity of the metric potentials across the boundary $r=R$ provides the following relationship :
\begin{eqnarray}
\left(1-\frac{2M}{R}\right)^{-1}&=&\frac{1 + 2aR^2}{(1 - bR^2)(1 + aR^2)},\label{o1}\\
1-\frac{2M}{R}&=& B^2(1 + aR^2),\label{o2}
\end{eqnarray}
and the pressure vanishes at the boundary, i.e., $p(r=R)=0$,
which provides the following equation:
\begin{eqnarray}
\frac{-4 b \pi + a \big\{4 \pi (1 - 5 b R^2) + \beta (3 - 4 b R^2)\big\}+
 2 a^2 R^2 \big\{4 \pi (1 - 3 b R^2) + \beta (2 - 3 b R^2)\}}{4 (\beta +
   2 \pi) (\beta + 4 \pi) (1 + 2 a R^2)^2}=0.\label{o3}
   \end{eqnarray}
\begin{itemize}

  \item {\bf Determination of $a,\,b$ and $B$ :}~ Solving the equations (\ref{o1})-(\ref{o3}) simultaneously, we obtain the expressions for $a,\,b$ and $B$ as,
  \begin{eqnarray*}
 a&=&\frac{8 M (\beta + 5 \pi) R^2 - (3 \beta + 8 \pi) R^3 +Q}{2 R^4 \{-12 M (\beta +
      4 \pi) + (5 \beta + 16 \pi) R\}},\\
      b&=&\frac{R^2 \{3 \beta R + 8 \pi (M + R)\} - Q}{4 (\beta + 4 \pi) R^5},\\
      B&=&\sqrt{\frac{R-2M}{R(1+aR^2)}},
 \end{eqnarray*}
The expression of the constant $Q$ is given by,
\begin{eqnarray*}
Q= \sqrt{
 R^4 \{\beta^2 (8 M - 3 R)^2 + 64 \pi^2 (M - R)^2 +
    16 \beta \pi (16 M^2 - 13 M R + 3 R^2)\}}.
 \end{eqnarray*}
 \end{itemize}
To calculate the numeric values of $a,\, b$ and $B$ the approximated mass and radius of the compact star LMC X-4 \cite{raw} is used and all the values are presented in Table~\ref{tb1} for different values of $\beta$.
\begin{figure}[htbp]
    \centering
        \includegraphics[scale=.55]{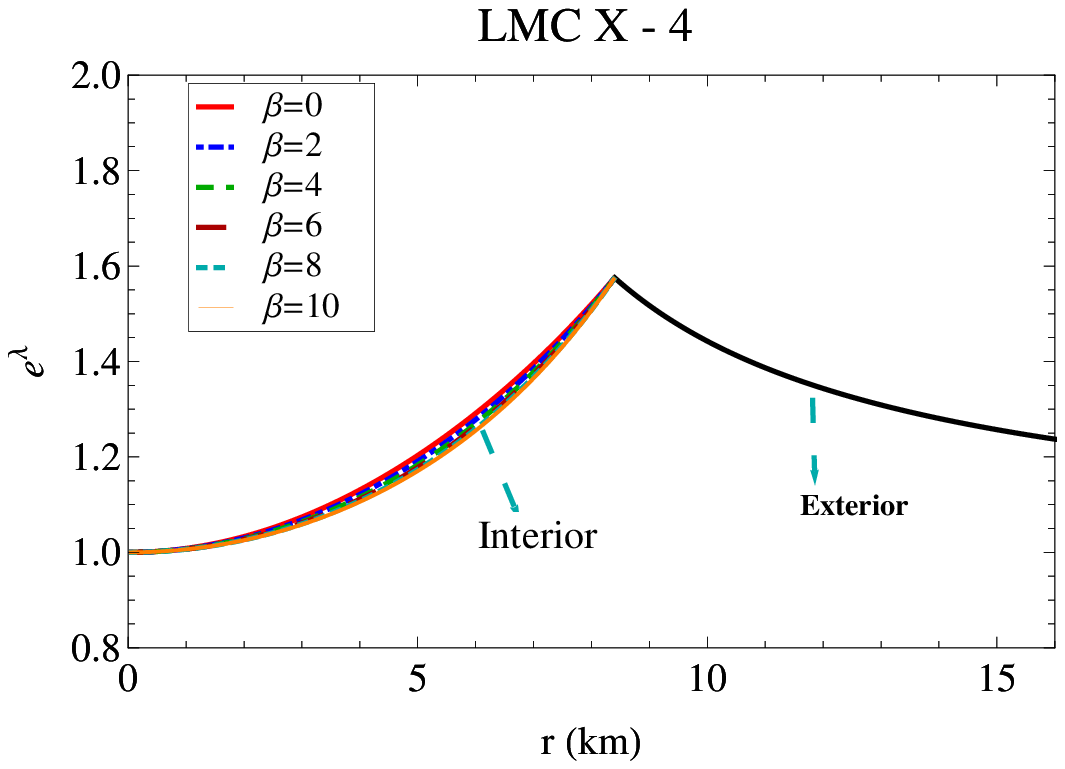}
        \includegraphics[scale=.55]{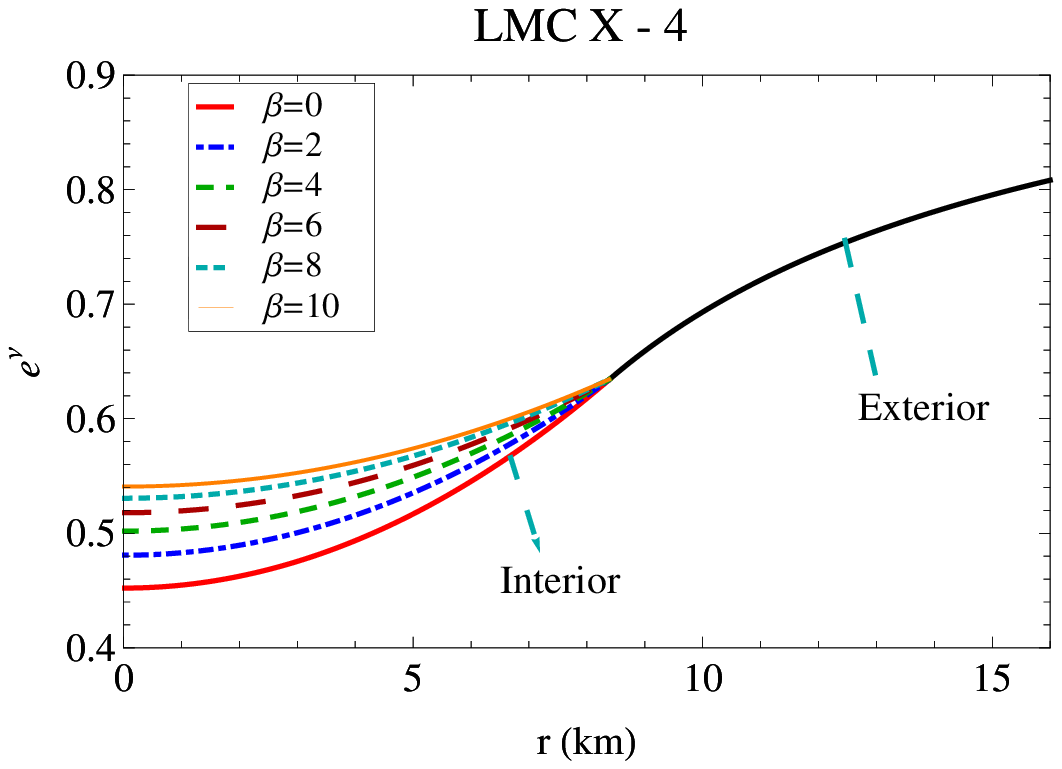}
       \caption{The metric coefficients $e^{\lambda}$ and $e^{\nu}$ are plotted against radius for different values of the coupling constant $\beta$ mentioned in the figure. The interior spacetime is matched to the exterior spacetime at the boundary. \label{metric}}
\end{figure}
\section{Physical analysis of the present model}\label{sec5}
In this section, we examine the matter density, pressure, mass function, compactification factor, analysis of redshift and adiabatic index etc. of our proposed models for different values of the coupling constant $\beta$.
\begin{itemize}
\item{{\bf Regularity of the metric coefficients :}}
At the center of the star, $e^{\lambda}=1$ and $e^{\nu}=B^2$,
  and their derivatives are given by, $(e^{\lambda})'= \frac{2 r \{b + a \big(1 + 2 b (r^2 + a r^4)\big)\}}{(1 + a r^2)^2 (1-b r^2)^2},\,(e^{\nu})'=2 a B^2 r$. The derivative of the metric coefficients vanishes at the centre of the star, implying that the metric coefficients are regular at the centre of the star. Fig.~\ref{metric} depicts the characteristics of metric coefficients.

\item{\bf{Nature of pressure and density and equation of state :}} The central values of density and pressure indicate that the solution is non-singular. The central values of these parameters can be obtained as,
\begin{eqnarray*}
\rho_c &=&\frac{4 b (\beta + 3 \pi) +
 a (5 \beta + 12 \pi)}{4 (\beta + 2 \pi) (\beta + 4 \pi)},\\
 p_c&=&\frac{3 a \beta + 4 a \pi - 4 b \pi}{4 (\beta + 2 \pi) (\beta + 4 \pi)},
\end{eqnarray*}
clearly both $\rho_c$ and $p_c$ are finite.
The equation of state describes the connection between pressure and density of matter. To model the compact object, many researchers employed linear, quadratic, polytropic, and other equations of state. To develop the stellar model in this study, we did not assume any specific equation of state. In Fig.~\ref{eos}, we have depicted the variation of pressure with respect to density using a graphical representation. The ratio of pressure to the density is also depicted in Fig.~\ref{eos} for different values of $\beta$.
    \begin{figure}[htbp]
    \centering
        \includegraphics[scale=.55]{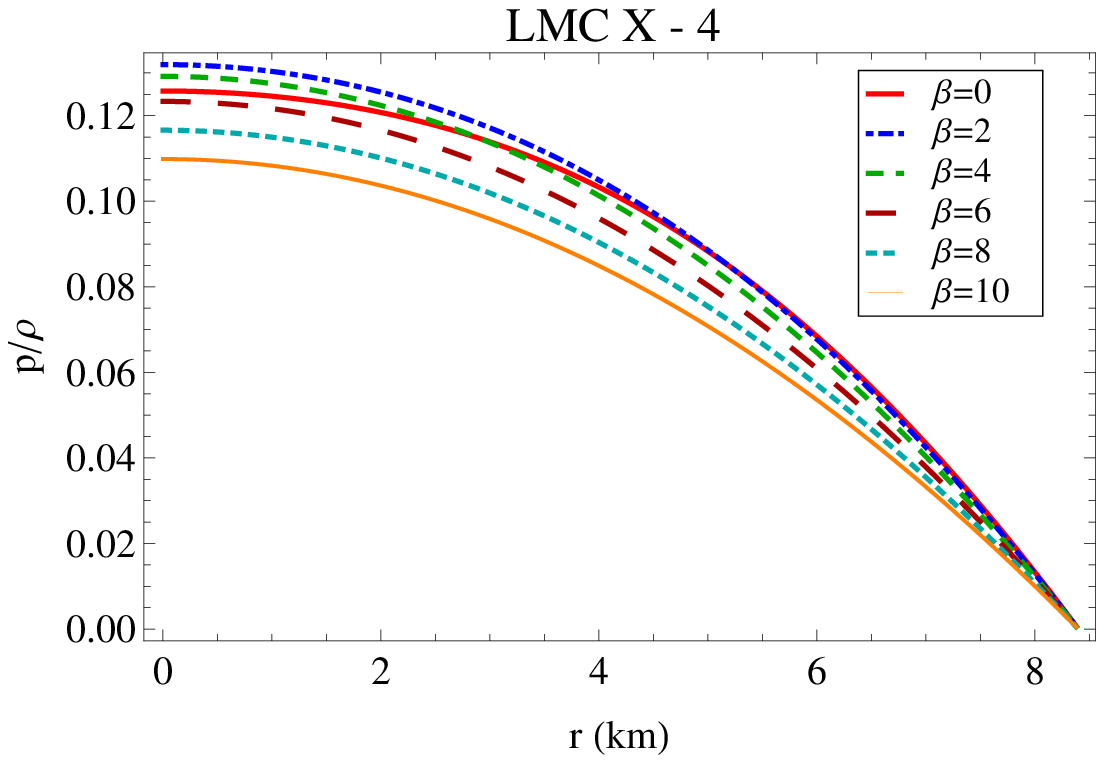}
         \includegraphics[scale=.55]{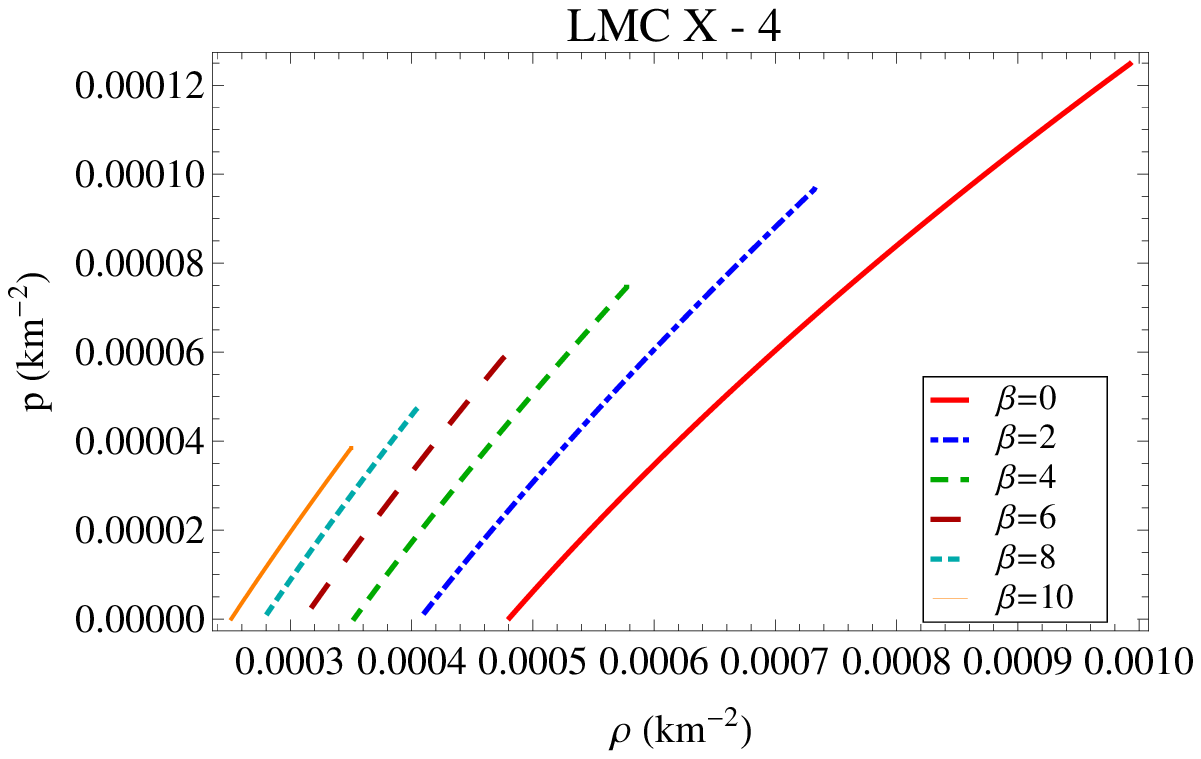}
       \caption{The ratio $p/\rho$ and the variation of pressure with respect to density are plotted against radius inside the stellar interior for different values of $\beta$.\label{eos}}
\end{figure}

\item{{\bf Relativistic adiabatic index :}}
For a given energy density, the stiffness of the equation of state can be characterized by the term adiabatic index, which also displays the stability of both relativistic and non-relativistic compact stars. Chandrasekhar \cite{70} introduced the concept of dynamical stability against infinitesimal radial adiabatic perturbation of the stellar system, and previous researches \cite{71,72} have successfully proven this hypothesis for both isotropic and anisotropic stellar objects. According to their estimations, the adiabatic index in all internal points of a dynamically stable stellar object must be greater than 4/3. For our present model, the expression of the adiabatic index is given by,
\begin{eqnarray}
\Gamma&=&\frac{4 (2 a + b) (\beta + 2 \pi) (1 + a r^2)}{3 a \beta + 4 a \pi -
 4 b \pi + 4 a \{a (\beta + 2 \pi) - b (\beta + 5 \pi)\} r^2 -
 6 a^2 b (\beta + 4 \pi) r^4}V^2
\end{eqnarray}
Fig.~\ref{sv} depicts the behavior of the adiabatic index $\Gamma$. The value of the adiabatic index is greater than $4/3$, as seen from the graph, confirming the stability of our proposed model.
\item{{\bf TOV Equation :}}
The hydrostatic equilibrium equation is an important attribute of the presented physical realistic compact object. By using the generalized Tolman-Oppenheimer-Volkov (TOV) equation, we can evaluate this equilibrium equation for our compact star candidate under the combined behavior of different forces. The generalized TOV equation for our present model in $f(R,\,T)$ modified gravity is given by,
\begin{eqnarray}\label{con1}
-\frac{\nu'}{2}(\rho+p)-\frac{dp}{dr}+\frac{\beta}{8\pi+2\beta}(p'-\rho')=0.
\end{eqnarray}
Under the combined action of three different forces, namely gravitational ($F_g$), hydrostatic ($F_h$), and the additional force due to modified gravity ($F_m$), the above equation predicts the stable configuration for the anisotropic celestial compact object.\\
\begin{figure}[htbp]
    \centering
        \includegraphics[scale=.55]{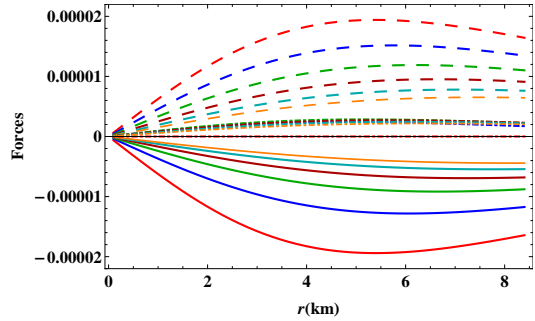}
       \caption{Different forces acting on the system are plotted against radius inside the stellar interior for different values of $\beta$. The description of the curves are as follows : The solid, large dashed and dot-dashed lines correspond to gravitational force, hydrostatics force and force due to modified gravity respectively. The color scheme is adopted as: red ($\beta= 0$), where $\beta=0$ corresponds
to GR, blue ($\beta=2$), green ($\beta=4$), darker red ($\beta=6$), cyan ($\beta=8$) and orange ($\beta=10$). \label{tov1}}
\end{figure}
The equation (\ref{con1}) can be written as,
\[F_g+F_h+F_m=0,\]
where,
\begin{eqnarray}
F_g&=&-\frac{a (2 a + b) r}{(\beta + 4 \pi) (1 + 2 a r^2)^2},\\
F_h&=&\frac{2 a (2 a + b) r \{\beta + \pi + a (\beta + 2 \pi) r^2\}}{(\beta +
   2\pi) (\beta + 4 \pi) (1 + 2 a r^2)^3},\\
   F_m&=&\frac{a (2 a + b) \beta r}{(\beta + 2 \pi) (\beta + 4 \pi) (1 + 2 a r^2)^3}.
   \end{eqnarray}
   The profiles of all the forces involved in the hydrostatic equilibrium condition are shown in Fig.~\ref{tov1}. The figure shows that the gravitational force counterbalances the combined behavior of hydrostatic force and modified gravity force, keeping our present system in stable equilibrium.
\item{{\bf Mass radius relationship :}} let us define compactification factor $\mathcal{U}$ as the following in terms of mass function,
\begin{eqnarray}
\mathcal{U}=\frac{\mathcal{M}}{R},
\end{eqnarray}
where $\mathcal{M}=m(r)|_{r=R}$.
The mass function $m(r)$ of the present stellar system is determined by,
\begin{eqnarray}
m(r)&=&4\pi\int_0^r \rho r^2 dr,\nonumber\\
&=&\frac{\pi }{16 a^{\frac{3}{2}}(\beta + 2 \pi) (\beta + 4 \pi) (1 + 2 a r^2)}\times\Big[2 \sqrt{a}
     r \Big\{(2 a + b) \beta + 4 a \big((4 a + 3 b) \beta + 8 (a + b) \pi\big) r^2 \nonumber\\&&+
      8 a^2 b (\beta + 4 \pi) r^4\Big\} -
   \sqrt{2} (2 a + b) \beta (1 + 2 a r^2) \tan^{-1}(\sqrt{2a}r)\Big]
 \end{eqnarray}
The mass of a compact star is directly proportional to its radius, as seen by the behavior of the mass function in Fig.~\ref{mass} for different values of $\beta$, indicating that mass is regular at the core. In this graph, we can see that the maximum mass is achieved at the boundary of the star.
Furthermore, the following formula can be used to calculate surface redshift ($z_s$) :
\begin{eqnarray}
z_s&=&\frac{1}{\sqrt{1-2\mathcal{U}}}-1,
\end{eqnarray}
The numerical values of compactification factor and surface redshift for different values of coupling constant $\beta$ are presented in Table.~\ref{tb2}.

\begin{figure}[htbp]
    \centering
        \includegraphics[scale=.55]{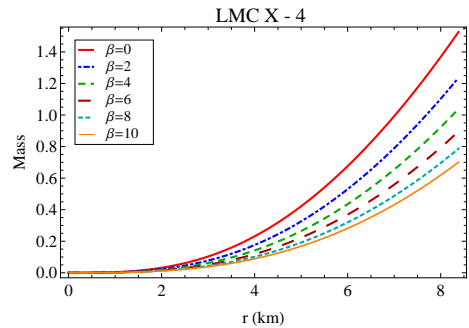}
       \caption{The mass function is plotted against radius inside the stellar interior for different values of $\beta$.\label{mass}}
\end{figure}

\end{itemize}

\begin{table*}[t]
\centering
\caption{The values of the constants $a,\,b$ and $B$ for the compact star LMC X-4 for different values of coupling constant $\beta$.}
\label{tb1}
\begin{tabular}{@{}cccccccccccccccc@{}}
\hline
Objects  & Estimated &Estimated & $\beta$ & $a$& $b$  &$B$ \\
&Mass ($M_{\odot}$)& Radius &&(km$^{-2}$)& (km$^{-2}$) &\\
\hline
LMC X-4  \cite{raw}& $1.04$&$8.4$& 0& $0.00572416$ & $0.00258814 $ & $0.672416$\\
&&& $2$ & $0.00453717$ & $0.00299468 $& $0.693418 $                                       \\
&&& $4$ & $0.00375275$& $0.00329288$& $0.708428$ \\
&&& $6$ & $0.00319696$ & $0.00352048 $& $0.719673 $ \\
&&& $8$ & $0.00278312$ & $0.00369964$ & $0.728402$ \\
&&& $10$ & $0.0024633$ & $0.0038442 $ & $0.735371$\\
\hline
\end{tabular}
\end{table*}

\begin{table*}[t]
\centering
\caption{The numerical values of central density, surface density, central pressure, compactness factor, surface redshift and central values of adiabatic index for the compact star LMC X-4 for different values of coupling constant $\beta$.}
\label{tb2}
\begin{tabular}{@{}cccccccccccccccc@{}}
\hline
$\beta$& $\rho_c$ & $\rho_s$ & $p_c$ & $\mathcal{U}$ & $z_s$& $\Gamma(r=0)$\\
\hline
 0& $1.33881 \times 10^{15}$& $6.47444 \times 10^{14}$ & $1.5153 \times 10^{35}$&0.182619&0.255147&1.79035\\
 2& $9.87685 \times 10^{14}$ & $5.48554 \times 10^{14}$& $1.17273 \times 10^{35}$&0.147996&0.191822&2.2384\\
 4& $7.78909 \times 10^{14}$ & $4.75465 \times 10^{14}$ & $9.05544 \times 10^{34}$&0.124365&0.153725&2.63319\\
 6& $6.41431 \times 10^{14}$ & $4.19343 \times 10^{14}$ & $7.11955 \times 10^{34}$&0.107202&0.128238&3.00513\\
 8& $5.44441 \times 10^{14}$ & $3.74943 \times 10^{14}$ & $5.71303 \times 10^{34}$&0.0941748&0.109981&3.36512\\
 10& $4.72527 \times 10^{14}$& $3.38966 \times 10^{14}$ & $4.67198 \times 10^{34}$&0.0839526&0.09626&3.71808\\
\hline
\end{tabular}
\end{table*}

\section{Discussion and Concluding Remarks}\label{sec6}

In this paper, we have investigated a model describing a relativistic fluid sphere in modified $f(R,\,T)$ gravity considering Tolman IV spacetime as interior geometry. Further, we have traditionally matched it with the exterior Schwarzschild geometry to evaluate unknown parameters present in our model.
To plot the different physical model parameters for our model we have considered the compact star LMC $X - 4$ with mass $(1.04 \pm 0.09)M_{\odot}$ and radius $8.301_{-0.2}^{+0.2}$ Km \cite{raw}. From the matching conditions and $p(r = R) = 0$ we obtain the values of $a$, $b$ and $B$.

LMC $X - 4$ is a X-ray pulsar discovered by the {\em Uhuru} observatory \cite{d1} and it is a high-mass binary system, which locates in the Large Magellanic Cloud(LMC) with estimated distance $d = 50$ kpc. The Large Magellanic Cloud (LMC) is a satellite galaxy orbiting our own, much bigger Milky Way. LMC $X - 4$ is a two-star system that includes a pulsar (a highly magnetized neutron star that beams X-rays) and a companion star \cite{d2} (and references therein). This system has approximately orbital period of $P_{orb}\simeq 1.4$ days \cite{d3}. In the earlier research works on LMC $X - 4$, Lee et al. \cite{d4} and White \cite{d5} have demonstrated the pulsar's eclipsing nature in X-rays, which is associated with a significant inclination of the system to the observer. The pulse period of LMC $X - 4$ is $13.5$ sec.

From our investigation we observe that, in the framework of $f(R,\,T)$ gravity as well, the Tolman IV spacetime acts similarly to the GR. The analytical equation of state has been derived from the metric. The physics of a compact star can be influenced by the universe's scale parameter because local interior spacetime of the compact star is a component of the global structure on the universal scale.
For the values of coupling constant $\beta = 0,\, 2,\, 4,\, 6,\, 8$ and $10$ the graphical illustrations of different physical properties have been presented in Figs.~\ref{pp}-\ref{mass} for the compact star LMC $X - 4$. The $\beta = 0$ corresponds to the GR.
Now these following points summarize our concluding remarks :
\begin{itemize}
\item  We obtain a clear picture of energy progression from left panel of Fig.~\ref{pp} for different values of $\beta$. The figure promises the real formation of stellar body having positive behavior inside the stellar interior and shows smooth declining nature towards the surface.
\item    The behavior of pressure $p$ in congruence with density, has been shown in right panel of Fig.~\ref{pp} for different values of $\beta$. This pressure plot declares the realistic formation of compact object as one can check that $p >0$ approaches to zero exactly at the boundary $r=R$. Also, the pressure is positive, continuous, maximum at the centre, then monotonically decreasing in nature, and do not suffer from any kind of singularities inside the stellar interior.
\item     The left and right panel of the Fig.~\ref{grad} represent the negative trend (Propagating from zero to negative from center to boundary) in gradient components of matter density and pressure.
\item      The profile of the both the metric potentials $e^{\lambda}$ and $e^{\nu}$ are plotted against radius for different values of the coupling constant $\beta$ in Fig.~\ref{metric}. A smooth matching of the metric potentials to the exterior spacetime at the boundary also has been shown in that figure.
\item    Our present model analysis shows that the null (NEC), weak (WEC), dominant (DEC) and strong (SEC) energy conditions are well satisfied throughout the stellar structure for our chosen $f(R,\,T)$ gravity model as shown graphically in the Fig. \ref{ec}. For the complexity of the expressions of matter density and pressure we have taken the help of graphical illustrations which ensures about the well behaved nature of all the energy conditions in $f(R,\,T)$ gravity.
\item Square of the sound speed (left panel) and relativistic adiabatic index (right panel) have been plotted versus radius in Fig. \ref{sv}. From the left panel of Fig. \ref{sv}, it is clear that the square of the sound speed lies within the predicted range throughout the fluid sphere i.e. $0<V^2<1$. Right panel of the Fig. \ref{sv} confirms the stability of our proposed model under adiabatic index $\Gamma>\frac{4}{3}$.
\item   The profiles of all the forces involved in the hydrostatic equilibrium condition are displayed in Fig. \ref{tov1}. The gravitational force $F_g$ counterbalances the combine effects of $F_h$ and $F_m$ to keep our proposed stellar model in equilibrium state as shown in the figure.
\item   We have described the ratio of pressure to the density versus radius (left panel) and variation of pressure with respect to density (right panel) using a graphical representation in Fig.~\ref{eos} for different values of $\beta$. One can note that $p/\rho$ lies in the range $(0,\,1)$ indicating the non-exotic nature of the matter distribution. Also we note that isotropic pressure $p$ obeys a linear relationship with matter density $\rho$.

\item The mass function is plotted against radius in Fig.~\ref{mass}. This figure shows that mass function is monotonic increasing function of radius and having no central singularity. The mass functional values are in agreement with required physical conditions as one can investigate from the figure.

\end{itemize}
From all obtained results and graphical illustrations, it is clear that our model is potentially stable and regular. Further detailed numerical features of our present model can be found in Table \ref{tb1} and \ref{tb2}. The numerical values of `a' decreases but `b' and `B' increase with increasing values of $\beta$. The central density $\rho_c$, surface density $\rho_s$, central pressure $p_c$, compactness factor and surface redshift all take lower values with increasing values of coupling parameter $\beta$. The central values of adiabatic index increase with increasing values of $\beta$, which concludes that the model becomes more stable for higher values of $\beta$.
Through analytical, graphical and numerical analysis, all the features of the model are well described. Finally, if we summarize our discussion, we collectively convinced by the calculated results which says that the system under discussion is physically reasonable and viably stable and our outcomes could be useful in modeling relativistic compact spherical objects in astrophysical phenomena such as quark and neutron stars.

\section*{Acknowledgements} P.B. is thankful to the Inter University Centre for Astronomy and Astrophysics (IUCAA), Government of India, for providing visiting associateship.


\begin{thebibliography}{99}

\bibitem{ob1} S. Perlmutter, et al., Supernova Cosmology Project collaboration, Astrophys. J. \textbf{517} (1999) 565.
\bibitem{ob2} C.L. Bennett, et al., Astrophys. J. Suppl. \textbf{148} (2003) 1.
\bibitem{ob3} A.G. Riess, et al., Supernova Search Team collaboration, Astron. J. \textbf{116} (1998) 1009.
\bibitem{ob4} P.A.R. Ade, et al., BICEP2 Collaboration, Phys. Rev. Lett. \textbf{112} (2014) 241101.
\bibitem{ob5} W.M. Wood-Vasey, et al., Astrophys. J. \textbf{666} (2007) 694.
\bibitem{ob6} M. Kowalski, et al., Supernova Cosmology Project Collaboration, Astrophys. J. \textbf{686} (2008) 749.
\bibitem{ob7} E. Komatsu, et al., WMAP Collaboration, Astrophys. J. Suppl. \textbf{180} (2009) 330.
\bibitem{ob8} M. Tegmark, et al., SDSS Collaboration, Phys. Rev. D \textbf{69} (2004) 103501.
\bibitem{ob9} K. Abazajian, et al., SDSS Collaboration, Astron. J. \textbf{128} (2004) 502.
\bibitem{ob10} E. Hawkins, et al., Mon. Not. R. Astron. Soc. \textbf{346}, 78 (2003) .
\bibitem{ob11} D.N. Spergel, et al., Wmap collaboration, Astrophys. J. Suppl. \textbf{148} (2003) 175.
\bibitem{fr1} S. M. Carroll, V. Duvvuri, M. Trodden, M. S. Turner, Phys. Rev. D \textbf{70} (2004) 043528.
\bibitem{fr2} S. Nojiri, S. D. Odintsov, Int. J. Geom. Methods Mod. Phys. \textbf{04} (2007) 115.
\bibitem{fr3} A.A. Starobinsky, Phys. Lett. B \textbf{91} (1980) 99.
\bibitem{fr4} A. Qadir, H.W. Lee, K.Y. Kim, Internat. J. Modern Phys. D \textbf{26} (2017) 1741001.

\bibitem{harko11} T. Harko, F.S.N. Lobo, S. Nojiri and S.D. Odintsov, Phys. Rev. D \textbf{84}, 024020 (2011)

\bibitem{frt5} V.U.M. Raoa, K.V.S. Sireesha, D.Ch. Papa Rao, Eur. Phys. J. Plus \textbf{129} (2014) 17.
\bibitem{frt6} B. Mishra, P.K. Sahoo, Sankarsan Tarai, Astrophys. Space Sci. \textbf{359} (2015) 15.
\bibitem{frt7} H. Shabani, Int. J. Mod. Phys. D \textbf{26} (2017) 1750120.
\bibitem{frt8} H. Shabani, A. H. Ziaie, Eur. Phys. J. C \textbf{77} (2017) 507.
\bibitem{frt9} M. Sharif, I. Nawazish, Eur. Phys. J. C \textbf{77} (2017)198 .
\bibitem{frt10} M. Sharif, M. Zubair, J. Cosmol. Astropart. Phys. \textbf{2012} (2012) 028.
\bibitem{frt11} M. Jamil, D. Momeni, M. Ratbay, Chinese Phys. Lett. \textbf{29} (2012) 109801.
\bibitem{frt12} M. E. S. Alves, P. H. R. S. Moraes, J. C. N. de Araujo, M. Malheiro, Phys. Rev. D \textbf{94} (2016) 024032.
\bibitem{frt13} Hina Azmat, M. Zubair, Eur. Phys. J. Plus \textbf{136}, 112(2021).
\bibitem{frt14} P.H.R.S. Moraes, J. D.V. Arbanil, M. Malheiro, J. Cosmol. Astropart. Phys. \textbf{2016} (2016) 06.
\bibitem{f1} P. Bhar, P. Rej, Eur. Phys. J. C {\bf 81}, 763 (2021)
\bibitem{f2}P. Bhar, P Rej, A Siddiqa, G Abbas, Int.J.Geom.Meth.Mod.Phys. {\bf 18} (2021) 10, 2150160
\bibitem{f3}P. Rej, P. Bhar and M. Govender, Eur. Phys. J. C {\bf 81}, 316 (2021).
\bibitem{f4} P. Rej,  P. Bhar, Astrophys Space Sci {\bf 366}, 35 (2021).
\bibitem{f5}P. Bhar, Eur. Phys. J. Plus {\bf 135}, 757 (2020).

\bibitem{frt15} D.R.K. Reddy, R.S. Kumar, Astrophys. Space Sci. \textbf{344} (2013) 253.
\bibitem{frt16} C.P. Singh, P. Kumar, Eur. Phys. J. C \textbf{74} (2014) 11.
\bibitem{frt17} H. Shabani, M. Farhoudi, Phys. Rev. D \textbf{90} (2014) 044031.
\bibitem{frt18} P.H.R.S. Moraes, Astrophys. Space Sci. \textbf{352} (2014) 273.
\bibitem{frt19} I. Noureen, M. Zubair, Eur. Phys. J. C \textbf{75} (2015) 353.
\bibitem{frt20} S.S. Yazadjiev, D.D. Doneva, K.D. Kokkotas, Phys. Rev. D \textbf{91} (2015) 084018.
\bibitem{frt21} M. Zubair, H. Azmat, I, Noureen, Int. J. Mod. Phys. D \textbf{27} (2018) 1750181.
\bibitem{frt22} I. Noureen, M. Zubair, Eur. Phys. J. C \textbf{75} (2015) 353.
\bibitem{frt23} M. Zubair, S. Waheed, Y. Ahmed, Eur. Phys. J. C \textbf{76} (2016) 444.
\bibitem{frt24} P. H. R. S. Moraes and P. K. Sahoo, Phys. Rev. D \textbf{97}, 024007 (2018).
\bibitem{frt25} M. Zubair, H. Azmat, I, Noureen, Int. J. Mod. Phys. D \textbf{27} (2018) 1750181.
\bibitem{frt26} M.J.S. Houndjo, Internat. J. Modern Phys. D \textbf{21} (2012) 1250003.
\bibitem{frt27} E.H. Baffou, A.V. Kpadonou, M.E. Rodrigues, M.J.S. Houndjo, J. Tossa, Astrophys. Space Sci. \textbf{356} (2014) 173.
\bibitem{frt28} G. Abbas, M. S. Khan, Zahid Ahmad, M. Zubair, Eur. Phys. J. C \textbf{77} (2017) 443.
\bibitem{frt29} G. Abbas, Riaz Ahmed, Eur. Phys. J. C \textbf{77} (2017) 441.
\bibitem{frt30} P.H.R.S. Moraes, J.D.V. Arbanil, M. Malheiro, J. Cosmol. Astropart. Phys. \textbf{06} (2016) 005.
\bibitem{hansraj} S. Hansraj, L. Moodly, Eur. Phys. J. C \textbf{80} (2020) 496.
\bibitem{c1} J. Lattimer (2010). http://stellarcollapse.org/nsmasses.
\bibitem{c2} R.H. Fowler, Mon. Not. R. Astron. Soc. \textbf{87} (1926) 114.
\bibitem{c3} S. Chandrasekhar, Mon. Not. R. Astron. Soc. \textbf{91} (1931) 456.
\bibitem{c4} S. Chandrasekhar, Mon. Not. R. Astron. Soc. \textbf{95} (1935) 207.
\bibitem{c5} M. Camenzind, Compact Objects in Astrophysics, Springer, Berlin, Germany, (2007).
\bibitem{c6} G. Abbas, A. Kanwal, M. Zubair,  Astrophys. Space Sci. \textbf{357} (2015) 109 .
\bibitem{c7} G. Fodor, arXiv:gr-qc/0011040.(2000)


\bibitem{c8} G. Abbas, Astrophys. Space Sci. \textbf{357} (2015) 158.
\bibitem{c9} M. Zubair, G. Abbas,  Astrophys. Space Sci. \textbf{361} (2016) 342.
\bibitem{c10} D. Deb, Rahaman, Farook, S. Ray, B.K. Guha, Phys. Rev. D \textbf{97} (2018) 084026.
\bibitem{c11} D. Deb, Rahaman, Farook, S. Ray, B.K. Guha, J. Cosmol. Astropart. Phys. \textbf{1803} (2018) 044.
\bibitem{c12} S. Hansraj, A. Banerjee, Phys. Rev. D \textbf{97} (2018) 104020.
\bibitem{c13} S. Hansraj, L. Moodly, Eur. Phys. J. C \textbf{80} (2020) 496.
\bibitem{c14} S.K.Maurya, A. Errehymy, Ksh. N. Singh, F. Tello-Ortiz, M. Daoud, Phys. Dark Uni. \textbf{30}, (2020) 100640.
\bibitem{c15} H. Azmat, M. Zubair, Eur. Phys. J. Plus \textbf{136} (2021) 112.

\bibitem{t1}B. Dayanandan, T.T. Smitha, Chin.J.Phys. {\bf 71} (2021) 683
\bibitem{t2}Ksh. N. Singh, S. Sarkar, F.Rahaman, Eur.Phys.J.Plus {\bf 135} (2020) 6, 484
\bibitem{t3}Estevez-Delgado, G., Estevez-Delgado, J., Soto-Espitia, R. et al. Eur. Phys. J. Plus {\bf 135}, 143 (2020)
\bibitem{t4}S. Banerjee, Pramana {\bf 91} (2018) 2, 27
\bibitem{t5}P. Bhar, Ksh.N. Singh, T. Manna, Astrophys. Space Sci. {\bf 361} (2016) 9, 284
\bibitem{t6}Ksh. N. Singh, N. Pradhan, N. Pant, Int. J. Theor. Phys. {\bf 54} (2015) 9, 3408
\bibitem{t7}M. Malaver, Int.J.Mod.Phys.Appl. {\bf 2} (2015) 1, 1
\bibitem{t8}J Ovalle , F Linares, Phys.Rev.D  {\bf 88} (2013) 10, 104026

\bibitem{landau} Landau, L.D., Lifshitz, E.M.: The Classical Theory of Fields. Pergamon,
Oxford (1998)
\bibitem{hi} T. Koivisto, Classical Quantum Gravity \textbf{23}, 4289 (2006).
\bibitem{raw}M. L. Rawls, J. A. Orosz, J. E. McClintock, M. A. P. Torres, C. D. Bailyn, and M. M. Buxton, Astrophys. J. {\bf 730}, 25
(2011).
\bibitem{pen}S.W. Hawking, G.F.R. Ellis, The Large Scale Structure of Space Time,
Cambridge University Press, Cambridge, 1975.
\bibitem{70} S. Chandrasekhar, Astrophys. J. 140 (1964) 417.
\bibitem{71} H. Heintzmann, W. Hillebrandt, Astron. Astrophys. {\bf 38} (1975) 51.
\bibitem{72} W. Hillebrandt, K.O. Steinmetz, Astron. Astrophys. {\bf 53} (1976) 283.



\bibitem{d1} R. Giacconi et al., Astrophys. J. \textbf{178}, 281 (1972).
\bibitem{d2} M. Falanga et al., Astron. Astrophys. \textbf{577}, A310 (2015).
\bibitem{d3} F. L. Lang et al., Astrophys. J. \textbf{246}, L21 (1981).
\bibitem{d4} F. Li et al., Nature \textbf{271}, 37 (1978).
\bibitem{d5} N. E. White, Nature \textbf{271}, 38 (1978).
\bibitem{tol}R.C. Tolman, Phys. Rev. 55, 364 (1939)

\end{thebibliography}

\end{document}